\documentclass[conference]{IEEEtran}
\usepackage[utf8]{inputenc}
\usepackage{color}
\usepackage{xfrac}
\usepackage{soul}
\usepackage{amssymb}
\usepackage[bookmarks, bookmarksopen, bookmarksnumbered, colorlinks,linkcolor=black,urlcolor=black,citecolor=black]{hyperref}
\usepackage{comment}

\usepackage{tikz}
\usetikzlibrary{shapes.geometric, arrows.meta, calc, positioning}

\tikzstyle{box} = [rectangle, thick, minimum width=3cm, minimum height=1.5cm, text centered, text width=6.5cm, draw=black]
\tikzstyle{arrow} = [thick, -Latex]
\tikzstyle{arrow2} = [thick, Latex-Latex]
\tikzstyle{arrowd} = [thick, dashed, -Latex]

\definecolor{green(html/cssgreen)}{rgb}{0.0, 0.5, 0.0}

\makeatletter
\newcommand\footnoteref[1]{\protected@xdef\@thefnmark{\ref{#1}}\@footnotemark}
\makeatother

\usepackage{fancyhdr}
\begin{document}

\title{Research Software Development \& Management in Universities: Case Studies from Manchester's RSDS Group, Illinois' NCSA, and Notre Dame's CRC}

\author{\IEEEauthorblockN{Daniel S. Katz\IEEEauthorrefmark{1}, Kenton McHenry\IEEEauthorrefmark{1}, Caleb Reinking\IEEEauthorrefmark{2}, and Robert Haines\IEEEauthorrefmark{3}} \IEEEauthorblockA{\IEEEauthorrefmark{1}National Center for Supercomputing Applications, 
University of Illinois at Urbana-Champaign, Urbana, IL, USA\\
Email: \{dskatz,mchenry\}@illinois.edu} \IEEEauthorblockA{\IEEEauthorrefmark{2}University of Notre Dame, Notre Dame, IN, USA\\
Email: caleb.m.reinking.2@nd.edu}
\IEEEauthorblockA{\IEEEauthorrefmark{3}University of Manchester, Manchester, UK\\
Email: robert.haines@manchester.ac.uk}}

\maketitle
\thispagestyle{fancy}

\begin{abstract}
Modern research in the sciences, engineering, humanities, and other fields depends on software, and specifically, research software. Much of this research software is developed in universities, by faculty, postdocs, students, and staff. In this paper, we focus on the role of university staff. We examine three different, independently-developed models under which these staff are organized and perform their work, and comparatively analyze these models and their consequences on the staff and on the software, considering how the different models support software engineering practices and processes.
This information can be used by software engineering researchers to understand the practices of such organizations and by universities who want to set up similar organizations and to better produce and maintain research software.
\end{abstract}

\section{Introduction}

Modern research in the sciences, engineering, humanities, and other fields depends on software, and specifically, research software. The NSF made 18,592 awards totaling \$9.6 billion to projects that mentioned ``software'' in their abstracts between 1995 and 2016~\cite{dia2}. An examination of 40 papers in Nature from January to March 2016 showed that 32 explicitly mentioned software, with each paper mentioning an average of 6.5 software tools, almost all of which were research software~\cite{nangia_nature}. Two surveys, of academics and Russell Group Universities in the UK~\cite{uk_survey} and members of the National Postdoctoral Association in the US~\cite{nangia_survey} found that about \sfrac{2}{3} of respondents said they couldn't do their research without software, while about \sfrac{1}{4} said they could, but it would be much more difficult, and only a few percent said it would make no difference.

Research software is most often produced by researchers themselves, often within academia,  by faculty, staff, postdocs, and students. While the academic environment and culture have developed over hundreds of years, software is much more recent. Many software projects are also developed and maintained over a longer time period than the academic tenure of any postdoc or student, and sometimes longer than staff and faculty as well. And until fairly recently, most software was developed in an ad hoc manner, without frameworks and tools (such as GitHub, science gateways, developer environments, software development kits, containers) to reduce the work to be done, and standard best (or even good enough) practices that provide researchers with effective methods for performing their software work, such as development and maintenance, so that research software can be shared.

A particular failing of the academic system in general is how it handles professional staff, and in particular, those staff who can understand research processes and methods, and who also understand good software practices. Such staff members might be members of a research group, hired by a faculty member and dependent on that person for their career and their livelihood, either formally as research staff or informally as a multi-year (or even multi-decade) postdoc. While these staff members may understand the research domain in which they work very well, and can support other shorter term staff (including research-oriented postdocs) and students, it can be difficult for them to learn about good software practices, except in a very \emph{ad hoc} manner. Alternatively, most universities have an IT department which has staff who have good software engineering expertise, but often don't understand how research is done, let alone have expertise in any particular research field.

In 2012, a small group in the UK attending the Software Sustainability Institute (SSI) Collaborations Workshop (\href{http://software.ac.uk/cw12}{http://software.ac.uk/cw12}) came together to address the question: why is there no career for software developers in academia?~\cite{rse_blog} The group members then wrote about their ideas on the people they were talking about, calling them Research Software Engineers (RSEs), their potential career paths, how their work is funded, and a set of recommendations later that year.~\cite{rse-dr12} The SSI's policy team then took up this challenge in 2013, which led community organizing, building the number of members of the UKRSE Association from 50 in 2013 to 633 in August 2016 to 1,272 in August 2018. These UK RSEs typically fit one of two different models: they are either embedded in research groups where they use a local RSE network to work together, or they are members of a central organization in their institution, where they work on various research projects on a short- or long-term basis. This RSE model has been discussed in the US as well, but without much university uptake so far, other than at Princeton.

This paper presents three different models for research software development: one ``traditional'' RSE group in the UK at the University of Manchester, and units in two US universities, specifically the Innovative Software and Data Analysis (ISDA, \href{http://ssa.ncsa.illinois.edu/isda}{ssa.ncsa.illinois.edu/isda}) group in the National Center for Supercomputing Applications (NCSA) at the University of Illinois at Urbana-Champaign, the Center for Research Computing Software Development Group (CRC, \href{https://crc.nd.edu}{crc.nd.edu}) at the University of Notre Dame. The Research Software and Data Science group at Manchester (RSDS, \href{http://itservices.manchester.ac.uk/research/services/software}{itservices.manchester.ac.uk/research/services/software}) currently comprises 23 RSEs. We describe the structure and operation of the Manchester group after a recent reorganization. ISDA comprises about 25 people, all of whom work at least partially on software. The CRC Software Development Group comprises about 24 people, 18 of whom work at least partially on developing software. Sections~\ref{sec:manchester}--\ref{sec:crc} describe the structure of these three organizations, how the organization impacts the software work that is done, and how the software work is actually done, including project management, code review, etc. Section~\ref{sec:comparison} compares these models and discusses software engineering lessons, and Section~\ref{sec:conclusions} concludes the paper.

\section{Manchester RSDS Model}
\label{sec:manchester}

The University of Manchester Research Software and Data Science (RSDS) group was founded in late 2014. As of March 2019, it currently comprises 23 RSEs. The group predominately works on short- to long-term RSE and Data Science projects (from weeks to years), but also supports the university's research applications---e.g., LabVIEW, MATLAB, Mathematica, etc.---and offers a wide range of related training, including Software Carpentry courses.


\subsection{Organizational context}

The group is based in Research IT, within the Directorate of IT Services. This ``central'' location---that is, outside of the three academic faculties---is important: it allows the group to work across all schools, departments, divisions, and institutes of the university with minimal administrative friction. Research IT was created as part of a wider transformation of the IT Services organization, but in reality this simply recognized and formalized research computing provision that had been going on within IT Services\footnote{And its previous incarnation, ``Manchester Computing''} for over 40 years.


\subsection{Research software engineering, data science, application support and training}

The original remit of the RSDS group was limited to providing research software engineering services to researchers and research projects on campus. These projects tended to last months or years, and typically involved a single RSE. The merger of the Research Applications Support group into the RSDS group meant that it also took on the shorter-term RSE work that group was doing, along with the support of research applications and training. All members now work on all aspects of the group's charge, working on projects, answering support tickets, and giving training where appropriate.

When the university's Data Science Institute was created, the model employed for providing ``software engineering as a service'' to research projects was also applied data science activities. Given that the RSDS group was already operating this model successfully, and that by their natures, an RSE and a data scientist are similar~\cite{rse_ds_common_blog}, the RSDS group was deemed a good fit for ``data science as a service'', too. 
%

\subsubsection{RSE projects}

An RSE project can be anything from a few days to a few years work, and might require anything from 0.2 Full Time Equivalent (FTE) and above of effort. In practice we have found that 0.2 FTE is about as low as it is possible to go and remain useful within a project. The largest project that the group currently works on comprises two FTE over ten years (and counting). RSEs on a research project can embed into the research group with which they are working. This can significantly improve the bootstrapping time for new projects, with the RSE able to assimilate a new domain area in a shorter time period. The collaboration between the research group and the RSDS group tends to bare fruit earlier and is more likely to result in more projects in the future. In this way the RSE also more easily becomes part of the ``institutional memory'' of the project; with RSEs on longer-term contracts than postdoctoral staff, they are more likely to be available for subsequent projects in the same domain.


With many current research projects so broad, and moving so quickly, flexibility is a unique selling point of the RSDS group. As a result of us maintaining a wide variety of skills in the team, and the fact that much of our engagement with projects is at around 0.5 FTE, we can easily move RSEs between projects should the need arise. This is particularly effective if the initial development on a project requires software engineering skills, with later analysis requiring data science skills, but it is equally useful where research takes an unplanned change of direction. In these cases we have, for example, been able to ``swap-out'' a mobile developer for a web developer at short notice when the software requirements of a project pivoted unexpectedly.

\subsubsection{Applications support}

The group provides second-line support for the university's major research applications and the most commonly used programming languages across campus. This includes validating and packaging installation media, helping people to optimize their code---either for their own hardware, or for the centrally provided HPC/HTC platforms---and short proof-of-concept work.

\subsubsection{Training}

Lastly, the group designs, develops and delivers training across a wide range of subjects, both to researchers across campus, and to external organizations. These training courses are mostly based on Software Carpentry~\cite{carpentries_web} materials, but extended and localized where appropriate. Although training is coordinated by the Applications Support and Training group within RSDS, we encourage any interested member of the group to undertake the Software Carpentry instructor training course and take part in training. We develop our own training materials using the Software Carpentry style and contribute them to the Carpentries if appropriate.

\subsection{Funding model}

The RSDS group is funded via three distinct routes:
projects funded by internal university research grants;
projects funded externally by the research councils, charities and industry;
and baseline funded\footnote{``Baseline funded'' here refers to work funded directly from overheads collected from fully-costed research grants by the central university.} activities.

For projects funded by internal university grants, where overheads such as estates costs\footnote{``estates costs'' is a catch-all to cover things like providing a computer, on a desk, in a building, that is cleaned and heated.} are not required, we use a ``day rate'' that covers the full cost of employing someone\footnote{On top of a person's salary, this includes taxes and pension contributions from the university.}. Externally funded projects are costed using full Economic Costing (fEC) rules to ensure that all costs of the university are covered. In both cases, the cost of each member of the team is amortized across all projects, rather than an individual being named for each project. This provides flexibility when assigning projects to each member of the group, without them needing to ``apply'' for each project as if it were a new job. It also allows for different RSEs to work on the same project with minimal administrative overhead. In this sense the RSDS group provides a pool of software engineers and data scientists, much like the other models discussed later.

Application support and training are baseline activities funded by the university. 
In all cases the salaries of the members of the group are underwritten by the IT Services organization. This means that continued employment is not reliant on specific grant income, so the team can be employed on permanent contracts, not fixed-term or so-called ``permanent-subject-to-funding'' contracts.

\subsection{Managing the group}

As a result of the rapid growth of the group, and the increasing reliance on the Head of Research Software Engineering to perform all acquisition of new business, all project management, and all line management of RSEs, the group was restructured in early 2019 with the aim of:

\begin{itemize}
  \item maintaining dedicated relationships between the group and important domain areas within the university---such as the faculties---and for cross-cutting specialisms---such as Data Science;
  \item a more consistent and up-to-date view of the incoming work pipeline and skills required in the group;
  \item more agile management of RSE projects ``in flight'';
  \item improved line-management of individuals within the RSE group.
\end{itemize}

\subsubsection{Group structure}

The RSDS group consists of: the Head of Research Software Engineering; six area leads; and a group of RSEs who work across all projects. The high-level leadership of the group is split into the following operational and domain areas: Applications Support and Training; Short RSE Projects; Data Science; the Faculty of Science and Engineering; the Faculty of Biology, Medicine and Health; and the Faculty of Humanities. The area leads are senior RSEs and data scientists within the group who spend 0.5 FTE on management activities and the rest on project work.


The rest of the RSDS group works in a ``matrix'' style; they each have a single line manager, but work on projects across all areas, so may have multiple project managers. In practice we find that RSEs are capable of managing themselves within projects, but that the project management support from the leadership group is important at certain times, and provides a useful point of contact for PIs. Consistent line management is important for the purposes of monitoring career development and training needs.

\subsubsection{Career paths}

People in the Manchester group all have a formal ``job description'' of Research Software Engineer, even if they might more closely identify as a Data Scientist.
Within the RSE job description are three levels, from junior to senior, that roughly equate to Research Assistant, Postdoctoral Research Associate and Lecturer on the university's academic track. This allows us to specify that new hires have adequate experience of research---for example a PhD or equivalent experience---while also being able to employ someone without, and develop them ourselves.

\subsection{Academic governance}

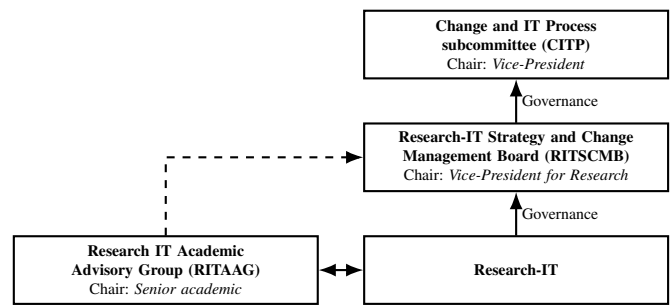
\begin{figure}[ht]
  \centering

  \begin{tikzpicture}[scale=0.6, transform shape, node distance=2.5cm]
    \node [box] (citp) {\textbf{Change and IT Process subcommittee (CITP)}\\Chair: \textit{Vice-President}};
    \node [box] (ritscmb) [below of=citp] {\textbf{Research-IT Strategy and Change Management Board (RITSCMB)}\\Chair: \textit{Vice-President for Research}};
    \node (rit) [box, below of=ritscmb] {\textbf{Research-IT}};
    \node [box] (ritaag) [node distance=1cm, left=of rit] {\textbf{Research IT Academic\\Advisory Group (RITAAG)}\\Chair: \textit{Senior academic}};

    \draw [arrow] (rit) -- node[anchor=west] {Governance} (ritscmb);
    \draw [arrow] (ritscmb) -- node[anchor=west] {Governance} (citp);
    \draw [arrow2] (ritaag) -- (rit);
    \draw [arrowd] (ritaag) |- (ritscmb);
  \end{tikzpicture}

  \caption{Research IT academic governance in Manchester
  \label{fig:man_r_it_gov}}
\end{figure}

Research IT's governance structure (see Figure~\ref{fig:man_r_it_gov}) fulfills two main purposes: it provides advice, ensuring that Research IT's work aligns with the university's strategic goals; and is a mechanism for approval and sign-off of internally funded projects. The role of the Research IT Academic Advisory Group (RITAAG) is primarily advisory; it is an open group, chaired by a senior academic, and is the primary mechanism by which Research IT can seek input from the academic community. 
The Research-IT Strategy and Change Management Board (RITSCMB) and Change and IT Process subcommittee (CITP) both provide official governance of Research IT activities. RITSCMB is composed of the faculty Vice-Deans for Research and other senior academics, and can approve projects with total costs lower than £250,000. CITP is a university-level committee that approves large projects that would have significant effects across the whole organization.

\section{Illinois NCSA Model}\label{sec:ncsa}
The NCSA Innovative Software and Data Analysis (ISDA) Group was founded roughly 15 years ago to address needs within the scientific community around image analysis and geospatial data support.  Serving to support scientific collaborators across a variety of domains, one mission of ISDA is to not only support a given effort's individual needs but to generalize those needs across projects and build software frameworks in response. The group then  maintains these frameworks and treats them as reusable artifacts that are sustainable beyond the lifespan of any one project. These frameworks in turn allow the group to engage new projects with overlapping and similar needs; the group can quickly ramp up such efforts with existing software in place, and within those projects, refine and extend the capabilities of the frameworks, which can then be leveraged by yet more future projects.  This paradigm of thinking about reusable software more than individual projects has not only allowed the group to survive what some  refer to as the ``funding Jenga'' of academic grants, with staggered grants of various sizes unpredictably coming and going over time, but  also to thrive, expanding from the group's three initial  full time research programmers/scientists members to 20+ across as many projects over a 10-year period.  The group has grown beyond its initial image and geospatial focus to support general software development, with projects ranging in size from supporting a quarter of a Full Time Employee (FTE) over a year to supporting 6+ FTEs over 5 years.

\subsection{Big Projects Anchor Small Projects, Small Projects Broaden Big Projects}
The ISDA group serves as a resource to scientific efforts, providing a pool of software developers with various backgrounds  along with a portfolio of tools.  The group further serves to offload and optimize the process of recruiting, retaining, training, growing, and managing staff, allowing these aspects to be minimized for the supported projects and their PIs.  A key aspect of sustaining this group of developers within an academic setting has been the careful management and alignment of supported efforts. 
Group leadership, described below, regularly reviews active and incoming projects and looks for opportunities to overlap efforts along a number of directions such as the software being leveraged and/or similarities in activities or domains supported by staff allocated to a project.
Given this, staff is then allocated so as to: amplify development activities, with tools and features being developed and/or leveraged across staffed projects; minimize the partitioning of staff and context switching by allowing team members to work on what is more or less a similar activity (though spread across different projects); maximize the ability for other team members to contribute to an effort when needed (e.g., a major demo or deliverable, staff members leaving the organization); creating and sustaining new frameworks by utilizing software created in one effort in other efforts (building a community around the software, perhaps across domains); and smooth over transitions in funding as projects end and new projects come in.
The ability align these efforts is much easier when one or two larger projects are part of the portfolio of projects being supported.  While not always possible (e.g., if differences in efforts  don't allow for overlap in activities), larger projects not only serve to reduce the juggling of staff from the management side but they also allow the group to more easily support much smaller projects, such as projects with smaller budgets on the order of 0.25 FTEs that couldn't support hiring a full developer. 
With staff members anchored to a related larger project, the smaller effort can be supported without stretching the staff member too thin (ISDA ideally tries to limit staff to two projects, three at the most) and developments within the smaller effort, which is also often shorter, can be sustained after the project ends.  As software sustainability is a concern for all software, whether small or large, the small projects help the larger effort by adding additional use cases and communities using the software.

\begin{figure*}[t]
\includegraphics[width=\textwidth]{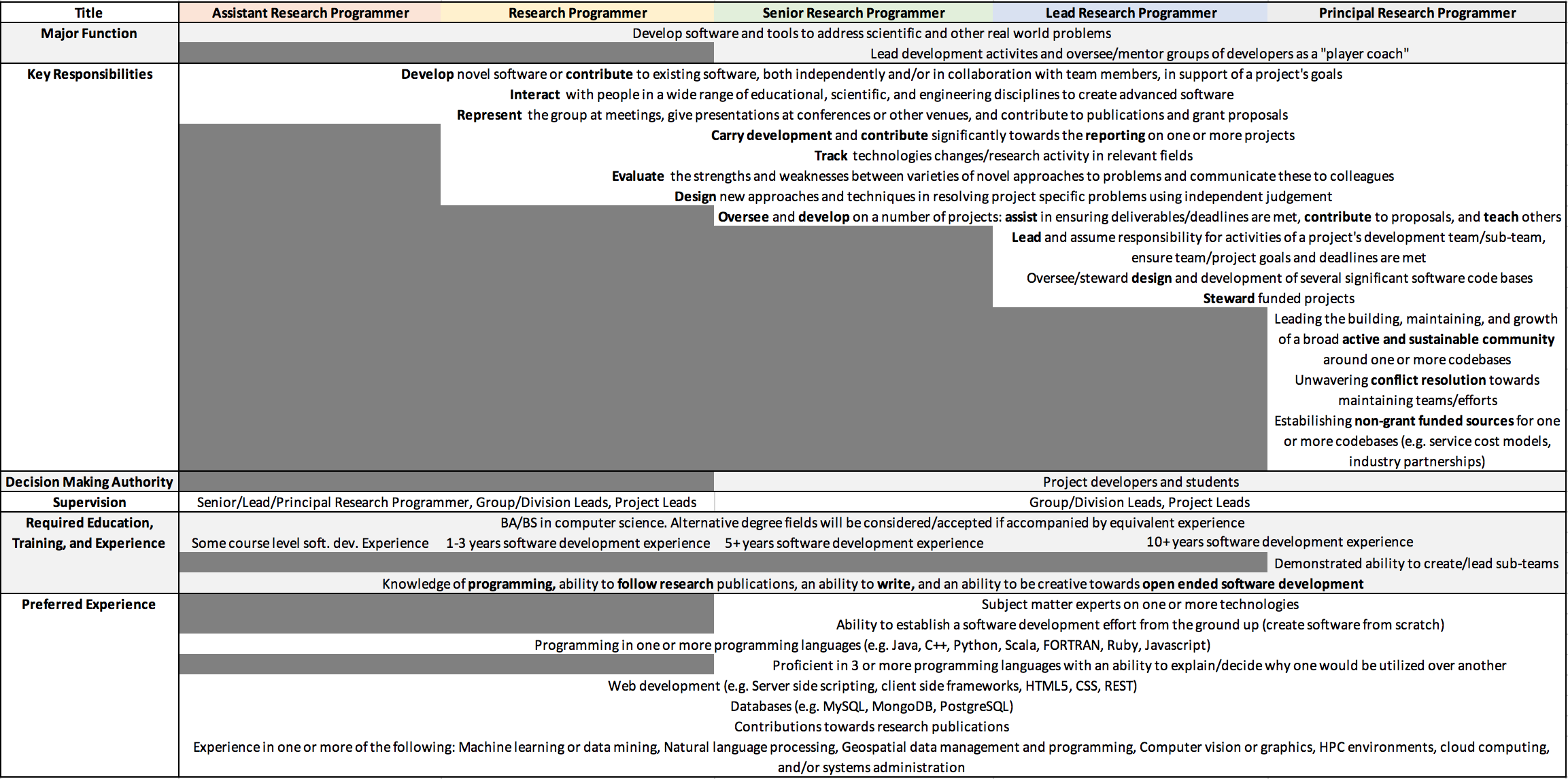}
\caption{Career path of research programmers at the National Center for Supercomputing Applications.
\label{fig:ncsa}}
\end{figure*}

\subsection{An Academic Model for Leadership and Career Paths}
For a variety of reasons, the group has evolved a leadership structure that in part mimics that of professors within a university setting, with a kind of ``latent'' tenure process as one advances through the ranks.  While there is no equivalent to the absolute job security of tenure due to the soft funding aspects of the organization, there is an increased capacity for job security as one advances along the research programmer career path, which rewards the creation of reusable software frameworks while simultaneously addressing the preservation of institutional knowledge.  
Specifically, new staff are mentored by senior staff in how to interact with scientific collaborators and students, and taught to think beyond the short term goals towards addressing longer term aspects such as reusability of developed software, maintainability, and overall sustainability.
Leveraging of existing software, possibly extending its capabilities, and/or the creation of new software that might seed a new reusable framework, is encouraged.  Research programmers who can do this become a resource within the group, in terms of supporting current/new frameworks across new projects and/or training staff/students on efforts utilizing that software.  Typically, these developers with one or more frameworks under their belt will find themselves with more projects to choose from than they could possibly be part of, requiring them to train and mentor newer research programmers themselves, in essence creating a sort of sub-team, in order to accommodate interested efforts.  This choice of projects around software they support provides the ``latent'' tenure and in turn increased job security, while simultaneously preserving the institutional knowledge in the codebases around those frameworks.  Research programmers following this path will be further mentored, if they choose, by senior developers in the development of proposals, serving as a Co-PI and/or PI, working with larger teams, management, and overall leading the technical/collaborative aspects of projects.

The group is lead by a Management Committee (MC) made up of the current body of senior developers overseeing the group's software frameworks and serving as PI/Co-PI/lead of the projects the group works on.  The MC in turn makes decisions about projects to take on, staffing assignments across the team, project priorities (in the sense of staffing increases, reallocation of senior developers for upcoming deliverables), as well as other financial/strategic decisions relevant to the group and the frameworks it oversees.
To grow the number of developers into this more strategic category, and to foster the ability to grow activities, the group has broken out the technical strategy aspects of the committee into a separate committee, the Technical Steering Committee (TSC), which serves to further mentor research programmers who have demonstrated an aptitude for advancing along the career path, doing so in a kind of ``hands-on'' approach by allowing such developers to take a larger part in strategic decisions across the projects served by the group (e.g., best underlying technologies to use, best practices, exposure strategies, etc.) and start taking on leadership roles within projects. 
The TSC further enhances the cross-pollination of activities and knowledge across the projects staffed by the group.  Each project staffed by the group has 1 month of a TSC/MC member, in addition to the development staff, to oversee, guide, and align efforts and serve as a backup in situations where additional development is needed or as the project memory in the event of the loss of a newer staff member.  The TSC is  about twice the size of the MC, roughly one up and coming research programmer to each current MC member.  Membership into these committees is based on nomination by an MC member followed by a vote.

The current career path for Research Programmers is depicted in Figure~\ref{fig:ncsa}.  Entry level positions deliberately have fairly minimal requirements, largely around an ability to program in one or more languages and a desire/ability to continuously learn.  This allows the group to be nimble in the recruitment of both newly graduated students who might otherwise move into industry and/or software developers that have a further interest in remaining in (or returning to) academia (either by being at the forefront of scientific discovery/research or possibly obtaining an advanced degree).  Each new level within the career path subsumes the skills of the previous level and adds additional skill requirements and responsibilities, in particular based on the idea of creating/leveraging reusable software, overseeing/pursuing projects, and mentoring newer developers.

\section{Notre Dame CRC Model}\label{sec:crc}
The Center for Research Computing at University of Notre Dame is an innovative and multidisciplinary research environment that supports collaboration to facilitate discoveries in science and engineering, the arts, humanities and social sciences, through advanced computation, data analysis and other digital research tools.
It enhances the university's cyberinfrastructure, provides support for interdisciplinary research and education, and conducts computational research.~\cite{crc_web}

\subsection{Organizational Context}
 The CRC, founded nearly 13 years ago with the primary service of maintaining computational resources for faculty researchers, is a part of the broader Notre Dame Research organization at the university. Since that time the center has added other distinct units, including specialized research faculty (formally titled Computational Scientists), the Center for Social Science Research, and the Software Development Group~\cite{crc_software_web}. The CRC currently employs nearly 60 staff.

\subsection{Agile Software Development in Research Context}
The CRC Software Development Group currently has about 20 staff members including software support positions and developers. Its role is to provide software development support and services to researchers. The group is broken up into four teams that all practice agile software development methods guided by the Scrum framework~\cite{agile_scrum}. Along with Research Programmers, each team also has a Product Owner (PO) and Scrum Master (SM, one per two teams). The Scrum Masters also serve as supervisors and team mentors for the staff on their teams. This has been interesting, as much of the Scrum literature, while not formally including guidance on management structure, predominately encourages a group to have supervision and management lines outside of the Scrum team. 

The CRC offers teams flexibility of process while still adhering to Scrum's foundational ideas of sprints, Scrum events, and building a potentially shippable product during every iteration. Code is hosted on GitHub under the CRC organization. Most teams currently use a two-week sprint, but other than this there are many differing practices within the teams. Differences include task/sprint management tools (TargetProcess~\cite{target_process_web}, JIRA~\cite{jira_web}, ZenHub~\cite{zenhub_web}, a whiteboard), how they work on concurrent projects (all working on same project in each sprint vs. working on all projects every sprint), and even the method of fulfilling the sprint events. The CRC gives each team a high level of empowerment to self-organize in these areas to create a productive environment.

When a prospective work opportunity is presented to the CRC, it follows a pattern of determining client needs and its ability to execute. This happens through a progressive set of meetings with the client. It is worth noting the a client can be either external to the CRC faculty from the Notre Dame campus, or internal CRC faculty who serves as a PI on a funded project. The first contact point is usually with the Assistant Director of Software Development (ADSD) who manages the overall CRC Software Development Group's portfolio of projects. This initial meeting helps to determine if the work that is needed fits into the mission/scope of the CRC's intended efforts on campus. If the project fits both the CRC's technical ability to deliver and the Center's mission, further meetings take place to define needs and desired outcomes, scope, and budget. Once these are understood, the ADSD evaluates team competencies and staff capacity and assigns the project to a team. After team assignment, the project will be handed off to that team and the associated Product Owner to manage and deliver the software. Much of the business development for new work is very organic and comes from repeat collaborations, internal collaboration with CRC computational scientists, or the CRC Director bringing opportunities to the software team through his collaborations with on and off campus groups.

Once work is accepted, the team assigned to the project works together to accomplish the project's desired outcome. It is the Product Owner's job to manage all projects within a team. This can range from 1 to 6 concurrent projects, based on size and duration of projects. A PO is responsible for managing client expectations, requirements, priorities, and budget on a project. They define requirements and objectives, mainly from the client's perspective, while deferring technical requirements and implementation techniques to the developers within the team. The PO closely and regularly communicates with the client, and can speak as the client within the team. The Scrum Master helps eliminate development or project management roadblocks to help the team achieve a peak efficiency and health. The SM is critical to creating a sustainable software development pace and cadence within the team. The Scrum Master can also look for ways to bring in subject matter experts to aid the team on difficult challenges.

Trying to share knowledge across the entire Software Development Group is a challenge.~\cite{knowledge-sharing} There is good technical support and sharing within each Scrum team, but challenges exist with sharing support and best practices across the organization. To attempt to provide a cross-team communication channel, the CRC has a Technical Leadership Team comprising the more senior developers from each Scrum team. This group meets regularly to share technologies and paradigms used within the teams. This group defines and carries out initiatives to improve the group's proficiency outside of the day-to-day project work. These include researching current topics, doing large group teaching and more one on one consultation with those who want it, and championing best practices within the team. The Technical Leadership Team has not been in existence long enough to make a decision on its effectiveness in facilitating this cross-team sharing and mutual benefit.

\subsection{Career Progression}
The CRC most closely resembles the NCSA model for programmers except it only encompasses the first three levels. The CRC's Research Programmer position falls under Notre Dame's general guidance of three distinct levels for individual contributor positions. The programmers within the CRC are mainly focused on applied software development and are not evaluated on publications or contributions to proposals. There is movement within the CRC to engage the programmers on more publications as a means to better understand the research process and to share their experiences with the broader academic community. This is an active area of evaluation for the CRC as clear career progression paths are identified as a strong need for staff retention. 

\subsection{Staff Funding Model}
The entire Software Development Group is funded through grants, contracts, or other collaborations and are not funded through hard funding lines. The CRC Research Programmers have all been hired on a permanent full time basis; there haven't been limited term contractors. 
This means that the CRC is generally conservative in hiring additional staff unless funding is certain for 6-8 months. In order to make decisions on opening new positions, the ADSD regularly forecasts staffing projections given the many existing and prospective projects that the center is working on. This is a difficult task due to the uncertainty of academic funding sources tied together with the urgency that comes once a project is funded. Based on this data, the CRC Director will make a decision on whether to open up a position. 

When working on projects, all time spent by Research Programmers is billed back to the faculty at an hourly rate. This hourly rate is calculated as a fully loaded rate including fringe, indirect costs, and some organizational overhead. On one hand, this funding model is transparent in the sense that faculty pay for actual work hours rather than approximate labor percentages. On the other hand, the addition of overhead into the rate creates challenges from the increased cost of programming resources compared to other staffing options such as students or postdoctoral staff.

\begin{table*}[ht]
    \centering
    \begin{tabular}{r|c|c|c|c|}
        \cline{2-5}
         & Effect & Manchester & Illinois & Notre Dame \\
        \cline{2-5}\cline{2-5}
        Institutional memory spanning projects, domains, time & $+$ & $\checkmark$ & $\checkmark$ & $\checkmark$ \\
        Flexible workforce with flexible skills & $+$ & $\checkmark$ & $\checkmark$ &           \\
        Can support varying levels of effort, in particular portions of staff members & $+$ & $\checkmark$ & $\checkmark$ & $\checkmark$ \\
        Supports mentoring/coaching & $+$ & $\checkmark$ & $\checkmark$ & $\checkmark$ \\
        Reduced bus factor with regards to project core knowledge & $+$ & $\checkmark$ &           & $\checkmark$ \\
        Enables scalable growth to more rapidly take on new/large efforts & $+$ &           &           & $\checkmark$ \\
        Fosters reuse and sustainability of built software & $+$ & $\checkmark$ & $\checkmark$ &           \\
        \cline{2-5}
        Costlier staff, however, better more maintainable code & $\pm$ & $\checkmark$ & $\checkmark$ & $\checkmark$ \\
        \cline{2-5}
        Perpetual precarious staffing allocations when solely reliant on grants & $-$ & $\checkmark$ & $\checkmark$ & $\checkmark$ \\
        Novelty and difference from status quo leads to difficulties describing to funding agencies/PIs & $-$ & $\checkmark$ & $\checkmark$ & $\checkmark$ \\
        Not possible to fund permanent staff under some agencies & $-$ & $\checkmark$ &           &           \\
        Risk of siloed staff after prolonged embedding in projects & $-$ & $\checkmark$ & $\checkmark$ &           \\
        Lack of assimilation into domain if project is too short & $-$ &           &           & $\checkmark$ \\
        Projects today too often do not consider/reward reuse & $-$ &           &           & $\checkmark$ \\
        \cline{2-5}
    \end{tabular}
    \caption{Comparison of discussed institutional RSE models via a set of  positive ($+$), mixed ($\pm$), and negative ($-$) aspects.
    \label{tab:comparison}}
\end{table*}

\section{Supporting Research Software Development}\label{sec:comparison}


The previous sections have described three different, organically evolving, organizational structures around the idea of supporting research software development with RSE-type staff members, and the means exercised by each organization to realize that idea. In this section, we draw attention to a number of similarities, or themes (depicted in Table~\ref{tab:comparison}) across these very disparate organizations.

\subsection{Institutional Memory Beyond a Single Project}

As software is becoming an ever more important, complex, and costly part of scientific research, it is increasingly recognized that better and longer-term leveragable software, as well as expertise, is valuable. Achieving this is difficult to do with students or support staff hired in an \textit{ad hoc} manner on a project by project basis.  As noted in \S\ref{sec:manchester}, RSE groups with longevity beyond any individual project, can act as ``institutional memory'' in these areas. RSEs who are employed in a central cross-cutting team typically have contracts that are independent of the lifecycle of particular research projects, so they may outlast other research staff employed on a particular project. When projects end, and post-docs leave to work elsewhere, the RSE remains in the pool, able to bring their experiences to bear in similar projects at a later date.  Further, RSEs who are generally more mobile across an institution than other research staff, and are more likely to have worked in a range of domains than those who focus on a career in a particular area, can spread knowledge/software artifacts obtained in one area over a wide number of users/communities. 

\subsection{Project/Team Management and Software Best Practices}

Research is increasingly a team endeavor.
As projects become larger and more complex, a wider range of skills is required.
Professionals who can write software and analyze data are becoming increasingly essential to many projects, where in many cases it is no longer acceptable to push these activities off to less experienced members of a research team, such as post-docs.
In turn, most projects that include RSEs still have a number of students, post-docs, and researchers involved in the research and development aspects.
RSE groups as a result tend to bring with them a more mature project management aspect/structure, provided through their senior developers and/or project managers.
These individuals bring with them notions of team organization (roles, responsibilities), work breakdowns and time-lines for deliverables (in addition to research goals), organizational tools (e.g., code repositories, issue trackers, wikis, instant messaging), and good coding practices (e.g., sprints, code reviews, integration testing and autobuilds), all of which lead to better, more sustainable code as well as more efficient group coordination.
These aspects are often barriers to many projects where the science questions within a particular domain are the main focus, and while anyone can read up on these practices and tools, having ready-to-go staff with  experience putting these aspects into practice has proven beneficial in terms of saving time adopting as well as successfully utilizing them.

\subsection{Overcoming Varying Finite Duration Funding Streams}
Attempting to retain long-term professional software development staff who work with scientists towards addressing research goals and are immersed in an academic research setting is challenging.
This must often be done with grant funding, which too often supports only part of such an RSE staff member, and are of short duration, making retaining such individuals long-term difficult if activities are not well coordinated to span many different projects over time.
Further, the dual leadership aspects of most projects supported by RSE groups can be confusing to software developers new to academia, with the Principal Investigators leading project activities and group leadership/managers driving group strength areas (e.g., web services, UI/UX) and addressing staff recruitment, development, retention, and career paths.
As a result, RSE groups seem to tend towards a form of ``Matrix Management,'' with one axis being project PIs over a set of funded projects that change over time, and the other axis being group managers who tend to be more fixed.
A challenge here is often in the management aspects, or overhead, typically not fundable by research grants.
Another aspect is in the need for a portion of the staff, at least senior staff, to set aside time and effort in the pursuit of new projects, e.g., grant writing.
This can be quite challenging without some core institutional funding source.

\subsection{Pushing to Change the Scientific Culture}
Scientific research will always be about scientific discovery first and foremost.
With a long and deep culture as to what this means (e.g., the scientific method, students, paper publications), adapting that culture to recognize the importance of newer technical aspects required by modern day science is very much an ongoing challenge.
These challenges span a number of areas, from constantly re-affirming the difference between RSEs and IT staff, to funding/budgeting differences (e.g., projects staffed by parts of multiple RSEs who may swap in/out over time as project needs change vs. permanent named individuals), to the need to elevate software itself to a first class citizen of scientific output and the need for software publication~\cite{software-citation}.  

RSEs and data scientists do not work in a vacuum; rather they are key to common research activities, such as hypothesis generation, study design, data analysis, and interpretation of results~\cite{code-theory-workshop-2017}. There are a number of efforts and communities addressing these issues~\cite{ReSA}; however, there is still effort needed to make these aspects well understood and accepted by the scientific community at large.

\section{Conclusions}\label{sec:conclusions}

With the growing importance of digital aspects that are increasingly recognized as necessary for modern day science, in particular software, RSE-type groups are emerging across organizations globally to address the needs around scientific software.  In this paper, we've presented three different, independently-evolved, organizational structures for research software development at universities.  While each has been developed at a different university, in the context of that university's particular characteristics, we believe that these models will be useful to others who are considering setting up such a group, and to those who already have a group and want to consider alternative structures that might improve the group's outcomes.
In addition, that fact that each model includes use of different software engineering practices means that these can be used as a data set for future scientific software engineering research, including the interaction between practices and processes and organizational structure.

Further, while software is becoming recognized as an essential part of research, many of the aspects necessary to support such software, e.g., RSE staff, RSE groups as a resource at the organizational level, and the need for better, more reusable software overall are less well-recognized.  
Additional topics that need more community discussion include recognition that an RSE is distinct from a typical industry programmer or a general academic IT staff member; how RSEs should be written into funding proposals; and how universities can provide support to address the non-grant fundable aspects of software development such as group management and long-term maintaining of software capabilities.

\bibliographystyle{ieeetr}
\bibliography{refs}

\end{document}